\documentclass[a4paper,english]{paper}
\usepackage[T1]{fontenc}
\usepackage[latin9]{inputenc}
\usepackage{listings}
\usepackage{float}
\usepackage{units}
\usepackage[numbers]{natbib}
\usepackage{subscript}

\makeatletter

\newcommand{\noun}[1]{\textsc{#1}}
\floatstyle{ruled}
\newfloat{algorithm}{tbp}{loa}
\providecommand{\algorithmname}{Algorithm}
\floatname{algorithm}{\protect\algorithmname}

\makeatother

\usepackage{babel}
\begin{document}

\title{Adaptive detection and severity level characterization algorithm
for Obstructive Sleep Apnea Hypopnea Syndrome (OSAHS) via oximetry
signal analysis%
\thanks{Ref.No: HG/BIOINF.0813.28v1 -- Last updated: 28-Aug-2013\protect \\
Preprint submitted to ArXiv.org -- http://arxiv.org/abs/...\protect \\
This work is licensed under a Creative Commons 3.0 License (BY-NC-SA)\protect \\
Copyright (c) by Harris V. Georgiou, 2013.%
}}

\author{Harris V. Georgiou%
\thanks{Email: xgeorgio@di.uoa.gr%
}}

\institution{Dept. of Informatics \& Telecommunications (MSc, PhD)\\
National Kapodistrian Univ. of Athens (NKUA/UoA), Greece}
\maketitle
\begin{abstract}
In this paper, an abstract definition and formal specification is
presented for the task of adaptive-threshold OSAHS events detection
and severity characterization. Specifically, a low-level pseudocode
is designed for the algorithm of raw oximetry signal pre-processing,
calculation of the 'drop' and 'rise' frames in the related time series,
detection of valid apnea/hypopnea events via SpO\textsubscript{2}
saturation level tracking, as well as calculation of corresponding
event rates for OSAHS severity characterization. The designed algorithm
can be used as the first module in a machine learning application
where these data can be used as inputs or encoded into higher-level
statistics (features) for pattern classifiers, in the context of computer-aided
or fully automated diagnosis of OSAHS and related pathologies.
\end{abstract}

\section{Introduction}

Obstructive sleep apnea-hypopnea syndrome (OSAHS) is a common disorder,
in which upper airway resistance is increased during sleep due to
upper airway dilator muscle relaxation and airway narrowing \citep{West2006}.
It is a common disorder and a recognized public health problem, affecting
roughly 2-4\% of adult male and 1-2\% adult female population \citep{Frederick1999,T.Young1993}.
It is still under-diagnosed and believed to be linked with severe
cardiovascular diseases, including hypertension, chronic fatigue,
metabolic disorders, daytime sleepiness, etc \citep{V.Kapur1999,SleepMedicineTaskForce1999,M.Partinen1989,J.He1988}.

Sleep studies for the diagnosis of OSAHS are performed are primarily
in a controlled environment, specifically via polysomnography (PSG),
where the patient is monitored during a full sleep cycle by various
electrophysiological signals, usually including electroencephalogram
(EEG), electro-oculogram (EOG) and electromyelogram (EMG), as well
as respiration and blood SpO\textsubscript{2} saturation tracking.
Moderate or severe OSAHS causes significant SpO\textsubscript{2}
desaturations in blood and usually this triggers the patient's awakening,
causing sleep fragmentation. When such events occur repeatedly, more
than five times per hour, it is considered a pathologically significant
state and the patient must undergo specific treatments. Since full
PSG is a difficult and tiresome procedure (the patient has to spent
the night in a sleep lab), oximetry-only monitoring via a non-intrusive
finger sensor is considered a very efficient and reliable, though
non-conclusive, means of detecting possible OSAHS pathology \citep{Frederick1999,F.delCampo2006,J.C.Vasquez2000,J.U.Magalang2003,W.FlemonsWard2003,Y.K.Lee2004}.
The rationale behind the use of oximetry-only OSAHS diagnosis relies
on the fact that, normally, the Apnea/Hypopnea Index (AHI) that is
calculated upon any respiratory discontinuities (events of typically
10 secs of cessation of air or longer periods of partial air flow
obstruction in the upper airway) during sleep is inherently correlated
to the SpO\textsubscript{2} level desaturation in the blood which
occurs almost immediately in such events due to hypoxia. Instead of
full respiratory tracking via PSG, the oximetry signal may be used
instead for the detection of such events \emph{by reference}. The
false-positive detections of such a procedure usually include other
pathologies that result in abnormal SpO\textsubscript{2} level fluctuations,
such as with the Cheyne-Stokes breathing, which also causes cyclic
SpO\textsubscript{2} level desaturations (e.g., heart failure, post
stroke) \citep{West2006}.

In this paper, an abstract definition and formal specification is
presented for the task of adaptive-threshold OSAHS events detection
and severity characterization. Specifically, a low-level pseudocode
is designed for the algorithm of raw oximetry signal pre-processing,
calculation of the 'drop' and 'rise' frames in the related time series,
detection of valid apnea/hypopnea events via SpO\textsubscript{2}
saturation level tracking, as well as calculation of corresponding
event rates for OSAHS severity characterization.

Algorithm \ref{alg:OSAHS-detector-specs} presents the typical definitions
for valid OSAHS events based on SpO\textsubscript{2} tracking, as
well as the corresponding severity levels based on events rate (per
hour) tracking. For the detection of potential OSAHS events, a common
clinical definition of a clinically-significant apnea/hypopnea event
for OSAHS diagnosis (based entirely on SpO\textsubscript{2} saturation
tracking) is employed \citep{West2006}. Specifically, this definition
is based on detecting a drop in oximetry level that is larger than
four points (-4\% SpO\textsubscript{2}) from current state, at \emph{any}
baseline value, within ten seconds or less. This essentially means
that \emph{any} dropping rate (negative gradient) in SpO\textsubscript{2}
saturation level of -24\% per minute or sharper may be tagged as a
potential apnea/hypopnea event. This approach is usually referred
to as \emph{continuous/adaptive threshold analysis} or \emph{moving
baseline} with regard to gradient calculations and comparisons \citep{A.Burgos2010}.
The justification behind this assertion is that \emph{any} such drop
in SpO\textsubscript{2}saturation, regardless of the starting level,
can not by attributed to normal SpO\textsubscript{2}fluctuations
during normal sleep when such events are regular (non-exceptions)
in the oximetry signal. 

Algorithm \ref{alg:OSAHS-detector-specs} also includes the typical
severity level scale for apnea/hypopnea event rates that is usually
applied for OSAHS diagnosis based on SpO\textsubscript{2} saturation
level tracking. Specifically, rare events at five or less per hour
are characterized as non-pathological, while more than five events
per hour are characterized as pathological at levels of increasing
OSAHS severity. Since the SpO\textsubscript{2} saturation level is
the only measurement (input) available in this framework, the detection
of any pathological OSAHS state (even 'mild') is usually a significant
(yet inconclusive) evidence in follow-up medical examinations, but
rarely the only basis for a final diagnosis of OSAHS. 

\begin{algorithm}
\caption{\label{alg:OSAHS-detector-specs}OSAHS events detector \& severity
levels (specifications)}

\begin{lstlisting}
\end{lstlisting}
\noun{if} (\emph{abs.drop of} $SpO_{2}$)$\geq$4 \noun{within} (\emph{timeframe})$\leq$10
\emph{sec} \noun{then} \emph{OSAHS event}$=$\noun{true}

\bigskip{}
OSAHS events rates \& severity levels: 
\begin{itemize}
\item R$\leq$5 \emph{(events/hour)} $\Rightarrow$ \emph{OSAHS severity}:
'\noun{normal}'
\item 5$<$R$\leq$15 \emph{(events/hour)} $\Rightarrow$ \emph{OSAHS severity}:
'\noun{mild}'
\item 15$<$R$\leq$30 \emph{(events/hour)} $\Rightarrow$ \emph{OSAHS severity}:
'\noun{moderate}'
\item R$>$30 \emph{(events/hour)} $\Rightarrow$ \emph{OSAHS severity}:
'\noun{severe}'\end{itemize}
\end{algorithm}

This paper describes the algorithmic description and proposed implementation
(as pseudocode) of the aforementioned definition and severity levels
for OSAHS. In the following sections, the overall algorithm for events
detector and event rates characterization is presented in a modular
way, first as a top-level outline of the processing queue and subsequently
each of the steps separately.

\section{Methods and Processing Queue}

Algorithm \ref{alg:OSAHS-detector-overview} presents the overall
processing queue of the oximetry-only signal with regard to OSAHS
events detection and severity characterization. The general definitions
in Algorithm \ref{alg:OSAHS-detector-specs} typically lead to gradient-based
methods for OSAHS events detection in the signal, usually after some
noise pre-filtering and null-value indices removal. 

Here, the processing pipeline presents a generic framework (in pseudocode)
for such a procedure, including all pre- and post-processing stages.
The pipeline includes five primary stages of signal processing in
a total of nine steps. In summary, steps 1-2 retrieve and pre-process
the oximetry signal by means of noise and missing-values removal via
specific low-pass filtering, i.e., without altering the low- and medium-frequency
statistics of the oximetry signal according to the OSAHS events detection,
as described by Algorithm \ref{alg:OSAHS-detector-specs}. Steps 3-4
create the corresponding oximetry gradient signal, while step 5 translates
this new data series to run-length histograms for further analysis.
Steps 6-7 processes the oximetry gradient signals and the corresponding
run-length histograms to mark detected OSAHS events, which are subsequently
timeframed and grouped according to their rates, i.e., the corresponding
OSAHS severity level. Finally, steps 8-9 perform the structured storage
of the output results and the final cleanup.

\begin{algorithm}
\caption{\label{alg:OSAHS-detector-overview}OSAHS detector -- Overview}

\begin{lstlisting}
\end{lstlisting}

\begin{enumerate}
\item retrieve raw SpO2 data series
\item pre-processing of raw data
\item stage-1: create SpO2 'state' (gradient sign) series
\item stage-2: patch any zeroes in the beginning of the 'state' series
\item stage-3: create rise/drop run-length histograms
\item stage-4: analyze runs and locate and OSAHS events
\item stage-5: analyze all the detected OSAHS events (rates)
\item store processed results
\item cleanup and exit\end{enumerate}
\end{algorithm}

The following sections describe each of these processing steps in
detail, with some remarks regarding to a possible implementation (as
real code).

\subsection{Pre-processing}

For the purposes of OSAHS events detection, the oximetry signal must
retain relatively smooth transitions between the samples, i.e., no
noise-related peaks or missing-value drops, while at the same time
preserve its original low-frequency characteristics where the most
important OSAHS-related informational content relies. 

The typical sampling rate of off-the-shelf oximetry sensors for long-term
monitoring is usually around 3 Hz (analog), which after rescaling
and some standard built-in local smoothing becomes a 'reliable' digitized
1 Hz data series of SpO\textsubscript{2} \% saturation level (70-100).
Hence, the specifications of Algorithm \ref{alg:OSAHS-detector-specs}
essentially refer to a sliding timeframe ('window') of at least 10
secs in length, tracking drops of SpO\textsubscript{2} saturation
level and marking as significant OSAHS event any gradient larger than
$\nicefrac{4}{10}$ or 0.4 Hz in the frequency range. Even at the
'reduced' sampling rate (digital) of 1 Hz the effective frequency
range spans up to 0.5 Hz, hence any such event should be clearly detectable
in the discrete-time signal after the built-in pre-processing of any
such typical oximetry sensor equipment. 

In this work, the proposed pre-processing steps are focused primarily
on removing any frequency elements higher than 0.4-0.5 Hz if a higher
sampling rate is used in the original signal (e.g., the raw 3 Hz analog),
as well as the removal of any missing/invalid values in the final
(digital) data series, which might still exist due to temporary faulty
measurement conditions. Normally, errors in the raw analog sampling
are corrected during the built-in digital-to-analog conversion process
(downsampling by smoothing), but if such errors span to more than
half the width of the smoothing kernel they will probably be detectable
in the digitized data series as 'gaps' (zero values) or invalid values
(e.g., negatives). These two functions, noise removal and missing-values
correction, may be implemented by a single low-pass digital filter
in the time domain (via a proper 1-D convolution kernel); however,
missing values may need some special detection and correction/replacement,
e.g., via localized linear interpolation, especially when they occur
in multiples, since they are not just 'negative peaks' in the signal
and can not be effectively removed by simple averaging (usually apply
median filtering or detection/replacement kernels).

The exact design and implementation of the pre-processing steps depend
heavily on the analog sampling rate, the new (downsampled) digital
sampling rate, as well as the quality and the noise properties of
the original oximetry signal, hence the equipment used is also an
important factor. In any case, at the end of the pre-processing steps,
the oximetry signal should be in the form of a properly filtered,
relatively noise-free ('smooth') data series, so that the corresponding
gradient series can be calculated reliably.

\subsection{Oximetry gradient sign series}

The first two stages of the core processing involves the calculation
of the oximetry gradient series, i.e., the change rates of the pre-processed
oximetry measurements against time. The calculation is performed in
two steps, namely the creation of the discrete differences and subsequently
a patching process for the correction of possible discontinuities
at the start of this new series.

Algorithm \ref{alg:OSAHS-detector-stage12} describes these two stages
in detail. Stage 1 translates the oximetry series into gradient sign
series (+/-) by employing a typical previous-value check in a sliding
window. In practice, OSAHS event detection do not require the analysis
of the exact gradient value but rather only its sign (rising or dropping),
as specified by Algorithm \ref{alg:OSAHS-detector-specs}. However,
the pseudocode in Algorithm \ref{alg:OSAHS-detector-stage12} can
be easily configured, if necessary, to store gradient values instead
of signs-only (see variable \emph{'change'} in line 3). Next, stage
2 back-patches any leading zeros at the start with the first non-zero
value that appears in the gradient series. This minor correction is
necessary for the next stage in the pipeline, i.e, introducing correct
run-length calculations (if employed) at the start of the oximetry
gradient series. 

\begin{algorithm}
\caption{\label{alg:OSAHS-detector-stage12}OSAHS detector, stages 1 \& 2 --
Calculate gradient series}

\begin{lstlisting}

stage-1: create SpO2 'state' (gradient sign) series
    for the entire SpO2 data series:
        calculate SpO2 change = current-previous
        if change>0 then mark SpO2 as 'rising':
            current='rise' , previous='rise'
        else if change<0 then mark SpO2 as 'dropping':
            current='drop' , previous='drop'
        else mark SpO2 as 'stable':
            current=previous
        end if
    end for

stage-2: patch any zeroes in the beginning of 
                           the 'state' series
    locate the first non-zero element
    backpatch elements up to the start
\end{lstlisting}
\end{algorithm}

It should be noted that, although the last checked condition in \ref{alg:OSAHS-detector-stage12}
is labeled as 'stable', no such state is registered; instead, the
previous definite state of 'rise' or 'drop' (+/-) is used. This is
because, according to standard OSAHS analysis and the specifications
in \ref{alg:OSAHS-detector-specs}, gradient sign reversal is strictly
defined. In practice, this means that a falling SpO\textsubscript{2}
saturation level that gets stabilized for a few samples is still considered
as falling, until a strictly positive gradient change is detected.
These strict detection conditions can be relaxed, if required, so
that non-changing oximetry values can be registered as separate 'stable'
states; however, this usually produces increased fragmentation of
the oximetry signal with regard to OSAHS events registration and,
hence, one pathologically significant 'long' OSAHS event (slow downward
trend) of gradually falling SpO\textsubscript{2} saturation level
may be mistakenly registered as multiple short 'insignifficant' events.
In the current framework, the detection and registration of OSAHS
events is considered within the 'strict' definition for gradient sign
changes, i.e., as described in \ref{alg:OSAHS-detector-stage12} above.

\subsection{Gradient sign run-lengths}

Stage 3 of the main processing pipeline involves the translation of
the oximetry gradient sign series into run-length statistics, so that
long runs that are relevant to real OSAHS events can be easily identified
and registered. Algorithm \ref{alg:OSAHS-detector-stage3} describes
this whole process in detail as pseudocode. 

As always, a maximum run-length size must be defined, which in this
case is set at 600 samples or 1 minute in real-time oximetry measurements
for a 1 Hz final (digitized, pre-processed) sampling rate as described
above. Using the already-calculated gradient sign series, the update
of the run-length matrix (RLM) is straight-forward and is completed
by a single run.

\begin{algorithm}
\caption{\label{alg:OSAHS-detector-stage3}OSAHS detector, stage 3 -- Create
RLM}

\begin{lstlisting}

stage-3: create rise/drop run-length histograms
    L = maximum run-length limit (typically 600)
    initialize run-length matrix (RLM) 2xL
    for the entire SpO2 'state' data series:
        calculate length of current run (up to limit L)
        characterize run as 'rise' or 'drop'
        update corresponding RLM cell
    end for
\end{lstlisting}
\end{algorithm}

As noted before, a 'strict' definition is applied with regard to gradient
sign changes, hence there are only two possible states, namely 'rise'
and 'drop' (i.e., no 'stable' state).

\subsection{Detection of potential OSAHS events}

The calculation and full update of the RLM statistics are in fact
not mandatory steps for the correct detection of OSAHS events; however,
these data contain valuable quantitative information about the oximetry
signal and its gradient and therefore they are usually involved in
the extraction of RLM-specific statistical features that can be later
analyzed and used as 'coders' for OSAHS pathological situations (e.g.,
input for pattern classifiers).

Algorithm \ref{alg:OSAHS-detector-stage4} describes the process of
detecting and registering possible OSAHS events in the oximetry gradient
sign ('state') signal \emph{without} the use of the RLM, as an example
of how an application with low computational-overhead can perform
this task. Also, this approach has the advantage of having no limits
on the exact length of the current run, as Algorithm \ref{alg:OSAHS-detector-stage3}
does with setting it to 600 samples (due to static RLM definition),
although this is usually a minor technical issue in practical software
implementations.

First, the gradient series is scanned and the current run limits are
calculated, and subsequently the identified timeframe is translated
into real time (secs). The full timeframe, state and limits are registered
and, if it is in fact an OSAHS event (see Algorithm \ref{alg:OSAHS-detector-specs}
specifications), it is marked as such for further processing. The
complete calculation of this stage is, again, a single-run processing. 

\begin{algorithm}
\caption{\label{alg:OSAHS-detector-stage4}OSAHS detector, stage 4 -- Mark
potential events}

\begin{lstlisting}

stage-4: analyze runs and locate and OSAHS events
    for the entire SpO2 'state' data series:
        calculate length of current run (no limit)
        locate the last position of the current run
        calculate timeframe length of the current run
        register current run (start,end,state,timeframe)
        if current run is OSAHS event (drop/time rule):
            register current OSAHS event data
        end if
    end for
\end{lstlisting}
\end{algorithm}

\subsection{OSAHS events rate and severity level}

As described earlier, the severity level of OSAHS is related to the
events \emph{rate} rather than their total sum during the monitoring
period. Therefore, it is necessary that all the detected events are
related to corresponding timeframes, i.e., the (maximum) number of
events detected within \emph{any} one-hour period during monitoring.

Using the results from the previous stage, i.e., Algorithm \ref{alg:OSAHS-detector-stage4},
the analysis of the event rates can be easily performed by examining
the corresponding registration data for each one of them. Specifically,
Algorithm \ref{alg:OSAHS-detector-stage5} analyzes the starting and
ending positions of each event, examines their placements within a
sliding timeframe of 60 minutes and calculates the total sum of occurrences
within these limits. There is also a timestamp correction for the
transition between 24-hour periods (from 11pm to 12am, i.e., the 23:59:59-to-00:00:00
entries reset). As the pseudocode describes, the process involves
the subsequent addition of the 'next' OSAHS event that is registered,
examining whether this is still within the current 60-minute time
window, and if not, removal of the 'last' OSAHS event and 'sliding'
the 60-minute frame forward. In other words, the one-hour window is
sliding \emph{event-wise} and not sample-wise, since all the OSAHS
events are already identified and registered during the previous processing
stages. This makes the calculation in stage 5 much faster and illustrates
how the event-based registration and the RLM (if present) can make
OSAHS-related analysis extremely efficient later on, possibly involving
the extraction of content-rich statistical features, with minimal
computational overhead. 

\begin{algorithm}
\caption{\label{alg:OSAHS-detector-stage5}OSAHS detector, stage 5 -- Calculate
event rates \& severity}

\begin{lstlisting}

stage-5: analyze all the detected OSAHS events (rates)
    W = OSAHS events per-hour rate window (lo/hi bounds)
    initialize the lower/upper bounds for W (both at 1)
    for the entire OSAHS events series:
        fix timeframe transitions between zones
                            (e.g., 11pm to 12am)
        update the lower/upper bounds for W
        if the ending time of current OSAHS event 
                           is still within 60 min
            increase the upper bound for W 
                            (expand frame)
            if rate within W > current maximum
                update max.rate frame in W
            end if
        else (if OSA event spans to more than 1 hour)
            increase lower bound for W (reduce frame)
        end if
        display OSAHS severity characterization 
                            (based on max.rate)
    end for
\end{lstlisting}
\end{algorithm}

It should be noted that, although the concurrent update and comparison
of two sliding windows (registered OSAHS events versus the 60-minute
frame) requires some delicate algorithmic formulation, the final pseudocode
is in fact of low computational complexity and very fast, since it
involves event-based and not sample-based calculations. This essentially
means that an oximetry series with minimal OSAHS events, i.e., 'normal'
cases, this stage may not introduce any significant computational
overhead at all.

\subsection{Post-processing}

At the finalizing steps, the application should store all the final
results and (some) intermediate calculation data for easy access.
No special post-processing is normally necessary here. If required
by the specific programming platform used by the exact software implementation,
any dynamic memory structures should be deallocated properly and any
open files should be buffer-flushed and closed here.

\section{Discussion}

As mentioned earlier, the purpose of this work is to present a low-level
specification of a complete OSAHS event detection algorithm, as well
as comments and suggestions regarding performance and reliability
issues. It is not limited to any specific software implementation
nor related datasets (benchmark or new); therefore, there are no full
experimental runs to be presented here.

Based on this low-level 'pseudocode' specifications presented in this
work, a prototype implementation has been developed in Matlab-compatible
code. Some benchmark oximetry/OSAHS datasets, as well as some custom
datasets (not available publicly), have been used for verification
and validation purposes%
\footnote{The Matlab-compatible implementation is still a work-in-progress,
currently in beta testing mode, developing enhancements for online
processing (see text for details), as well as cross-datasets compatibility.
When finalized, it will be made publicly available for download (on
request) from the author's website.%
}.

\subsection{Technical issues}

One of the most important items of this framework for the correct
detection and labeling of OSAHS events is the correct calculation
of the oximetry gradient series. Algorithm \ref{alg:OSAHS-detector-stage12}
describes this calculation as a simple difference between the current
and the previous value, i.e., using a 2-value wide difference operator.
However, the gradient values may be calculated by employing a wider
kernel, i.e., an operator with width larger than 2 (e.g., a 3-value
centralized mask), in order to better compensate with any remaining
noise artifacts and/or improve the relation to the \emph{momentum}
(2nd-order properties) of the gradient rather than its spot value.
Additionally, the specific limits for SpO\textsubscript{2} drop rate
(-4\%) and the corresponding time frame ($\leq$10 seconds) can be
adjusted to more strict or more relaxed values, based on the required
sensitivity/specificity of the OSAHS event detector.

The detailed description of all the steps in this framework, as outlined
by Algorithm \ref{alg:OSAHS-detector-overview}, is based on the continuous
tracking of the of SpO\textsubscript{2} saturation level as registered
in the oximetry data series. This approach is usually referred to
as \emph{continuous/adaptive threshold analysis} or \emph{moving baseline}
with regard to gradient calculations and comparisons \citep{A.Burgos2010}.
Other approaches include \emph{multi-threshold} analysis of the oximetry
signal \citep{A.Burgos2010,N.Oliver2007}, where there are several
pre-defined levels of SpO\textsubscript{2} desaturation and each
'drop' state is characterized by the appropriate \emph{desaturation
index} for the time frames it remains below every such level, namely
ODI4 (Oxygen Desaturation Level) for -4\%, ODI3 for -3\%, etc. Furthermore,
the time spent below any pre-defined desaturation level can also be
characterized by appropriate indices, namely the TSA90 (Time Spent
in Apnea) for a 90\% threshold, TSA88 for a 88\% threshold, etc. All
these indices can be embedded in the processing stages described later
on, as they rely on simple threshold checks. 

These potential events can be analyzed subsequently by other filtering
factors, e.g., multiple drop rate levels, and labeled as 'true' or
'insignificant' apnea/hypopnea events for OSAHS diagnosis. However,
since these are referred to the statistical characterization and \emph{coding}
of the oximetry signal into specific markers of \emph{features} that
can be used as inputs in pattern classifiers, this type of analysis
is not considered in this work, which is considered only with the
detection, registration and characterization of OSAHS events and event
rates (severity level).

\subsection{Enhancements \& extensions}

The detailed description of all the steps in Algorithm \ref{alg:OSAHS-detector-overview}
is based thus far on the assumption of batch or 'offline' processing:
the entire oximetry data series is assumed complete and available
at full length before the processing begins. Strictly speaking, this
requirement is not necessary for the detection of OSAHS events and
it is employed here for technical reasons only, since it makes the
design and description of all the intermediate steps much simpler
and straightforward. In practice, having the entire data series available
from the start simplifies its scanning when producing the corresponding
gradient series, the run-length matrix (RLM), as well as the proper
registration of every OSAHS event detected. However, this requirement
can be easily lifted in two ways, namely: (a) by changing the intermediate
processing steps of the pipeline as to work with dynamic limits and
thresholds that are updated adaptively as the data series is being
generated, or (b) by using the presented framework in a \emph{localized}
way, i.e., processing the data series locally by using a proper \emph{sliding
window} technique.

With regard to the first choice, i.e., implementing a fully adaptive
'online' version of the framework, there are some comments and suggestions
that should be taken into account:
\begin{itemize}
\item Algorithm \ref{alg:OSAHS-detector-stage12} (stages 1 \& 2): No major
changes are needed here. In stage 2, back-patching the gradient series
to the beginning is now unnecessary, since the oximetry data series
is being generated on-the-fly, so is the corresponding gradient series.
\item Algorithm \ref{alg:OSAHS-detector-stage3} (stage 3): Here, the RLM
calculation must be converted to a \emph{running} RLM, since the processing
pipeline is now dynamic. In practice, this means that all RLM updates
should be made on-the-fly as soon as an OSAHS event is registered
as 'ended'. Normally, this requires a new set of RLM variables, possibly
not a new 'temporary' RLM, in order to update the current run as the
data are being generated and then flushed to the proper RLM entry
for global update.
\item Algorithm \ref{alg:OSAHS-detector-stage4}: As in the case of RLM
calculations, this stage should also be implemented as a \emph{running}
OSAHS event detector. This means that event start, tracking and ending
should be calculated and registered on-the-fly, essentially using
an additional set of limit and threshold variables as the ones presented
for the 'offline' version of this framework. The application of the
60-minute timeframe for OSAHS rate calculation is now completely straightforward,
as it always examines only the most recent 60 data samples, but each
OSAHS event must be kept in a temporary record first before its details
are fully determined and can be properly registered in the global
record. Normally, the most recent SpO\textsubscript{2} value should
always be considered as a possible 'start' (if none is active) or
'end' (if one is already active) of an OSAHS event, which means that
the corresponding event window must be kept 'open' and dynamically
updated during the data generation process. This requirement may have
negative effects on the quality of the event detections (false positives/negatives)
if similarly dynamic pre-processing is not employed on-the-fly (simple
pre-processing as described in Algorithm \ref{alg:OSAHS-detector-overview}
might not work).
\item Algorithm \ref{alg:OSAHS-detector-stage5}: After every new OSAHS
event is detected, identified as 'ended' and properly registered,
stage 5 should be triggered in order to examine it immediately for
determining its severity level. In other words, the OSAHS events are
not examined after they are all fully registered as in the 'offline'
version but now each one is examined and labeled as soon as its boundaries
(start, end) are determined.
\end{itemize}
The second choice for creating an 'online' version of the OSAHS detector
is, as mentioned previously, the employment of an active timeframe
or sliding window technique. In this case, the framework is applied
as-is, but in a localized fashion: the oximetry data series is processed
in overlapping blocks, large enough for any expected OSAHS event (e.g.,
one full 60-minute block) and global registries, as well as OSAHS
rates and severity level, are updated only when necessary, i.e., when
new events are registered. This approach has the advantage of keeping
the overall algorithmic complexity low, exactly as the original version
of this framework, while at the same time limiting the required resources
to the absolute minimum, e.g., memory buffers for only 60-minute timeframes
of data.

Both cases, the 'windowed online' version and the 'fully dynamic online'
version, are highly parallelizable. If necessary, stages 1 through
5 can be easily implemented as separate threads or tasks in a multiprocessing
environment, as long as there are proper synchronization mechanisms
between them. Of course the overall processing is still a pipeline,
i.e., inherently sequential, but its subsequent stages can be implemented
in a highly overlapping fashion, especially between stages 1 and 2
(main data series processing), and between stages 3 to 5 (RLM updates
and events registration/rates/severity).

\section{Conclusion}

In this paper, an abstract definition and formal specification was
presented for the task of adaptive-threshold OSAHS events detection
and severity characterization. Specifically, a low-level pseudocode
was designed for the algorithm of raw oximetry signal pre-processing,
calculation of the 'drop' and 'rise' frames in the related time series,
detection of valid apnea/hypopnea events via SpO\textsubscript{2}
saturation level tracking, as well as calculation of corresponding
event rates for OSAHS severity characterization.

The designed algorithm covers the preliminary phase of coding in oximetry
signal analysis, i.e., the detection, registration and characterization
of all OSAHS events with regard to their bounds, rates and OSAHS severity
level. Therefore, it can be used as the first module in a machine
learning application where these data can be used as inputs or encoded
into higher-level statistics (features) for pattern classifiers, in
the context of computer-aided or fully automated diagnosis of OSAHS
and related pathologies.

This framework was considered under a standard 'offline' (batch) version
of processing, as well as possible enhancements for 'windowed online',
'fully online' and parallelizable versions with minimal requirements
with regard to memory resources (e.g., for implementations on embedded/mobile
devices).

\subsubsection*{Acknowledgments}

{\small{The author wishes to thank Mr. Theodoros G. Papaioannou, assistant
professor in Biomedical Engineering at the Medical School of National
Kapodistrian University of Athens (NKUA/UoA), and Mr. Eleftherios
Kosmas, sleep technologist at National Kapodistrian University of
Athens (NKUA/UoA), for their collaboration during the preliminary
stages of this work, with regard to the exact specifications of OSAHS
events and for providing some test cases (prototype datasets) for
testing purposes of the early software implementations.}}{\small \par}

\bibliographystyle{plain}
\nocite{*}
\bibliography{osahs-detector-algorithm_paper_ver1}

\end{document}